\documentclass[a4paper]{jpconf}

\usepackage{graphicx}
\usepackage{amsmath,amssymb}
\usepackage{multirow}
\usepackage{wrapfig}

\graphicspath{{figures/}}

\begin{document}

\title{Late time cosmology with LISA:\\ Probing the cosmic expansion with massive black hole binary mergers as standard sirens}

\author{Nicola Tamanini}

\address{Institut de Physique Th\'eorique, CEA-Saclay, CNRS UMR 3681, Universit\'e Paris-Saclay, F-91191 Gif-sur-Yvette, France}

\ead{nicola.tamanini@cea.fr}

\begin{abstract}
	This paper summarises the potential of the LISA mission to constrain the expansion history of the universe using massive black hole binary mergers as gravitational wave standard sirens.
	After briefly reviewing the concept of standard siren, the analysis and methodologies of Ref.~\cite{Tamanini:2016zlh} are briefly outlined to show how LISA can be used as a cosmological probe, while a selection of results taken from Refs.~\cite{Tamanini:2016zlh,Caprini:2016qxs} is presented in order to estimate the power of LISA in constraining cosmological parameters.
\end{abstract}
 
\section{Introduction}

The LISA mission \cite{elisaweb} aims at measuring gravitational waves (GWs) in the frequency band around the mHz.
Among the richness of astrophysical sources expected to produce a detectable GW signal at those frequencies, there are massive black hole binaries (MBHBs) from $10^4$ to $10^7$ solar masses.
The inspiral, merger and ringdown of several MBHBs will be observed by LISA with an high signal to noise ratio (SNR), allowing for an accurate estimation of the parameters of the binary.
The information gathered from these GW sources will lead to advances not only in astrophysics, but also in cosmology.
In fact MBHBs can be used as reliable distance indicators since the luminosity distance to the source is one of the parameters entering the measured waveform.
The joint observation of a GW signal and an electromagnetic (EM) counterpart, from which the redshift of the source can be extracted, will thus allow to test the cosmic expansion through the distance-redshift relation, in a similar fashion to type-Ia supernovae (SNIa) analyses.
In what follows, after briefly reviewing how a compact binary system can be used to constrain the cosmic evolution, the analysis and techniques developed in \cite{Tamanini:2016zlh} will be outlined, and a selection of results from \cite{Tamanini:2016zlh,Caprini:2016qxs} regarding forecasts for LISA will be presented.

\section{The concept of standard siren}

At the lowest (Newtonian) order, the waveform produced by a binary astrophysical system and observed by an interferometric detector on Earth or in space, can be parametrized as (see e.g.~\cite{michele})
\begin{align}
	h_\times = \frac{4}{d_L(z)} \left( \frac{G \mathcal{M}_c(z)}{c^2} \right)^{5/3} \left( \frac{\pi f}{c} \right)^{2/3} \cos\iota \sin[\Phi(f)] \,,\label{eq:gwf}
\end{align}
with a similar expression holding for the $h_+$ polarization, where a $(1+\cos^2 \iota)/2$ dependence on $\iota$, the angle characterizing the orientation of the binary orbital plane, is present.
Here $\mathcal{M}_c(z)$ is the chirp mass of the binary, $d_L(z)$ is the luminosity distance, $f$ is the GW frequency at the observer and $\Phi(f)$ is the phase of the GW.
As pointed out for the first time by Schutz \cite{Schutz:1986gp}, the parameter estimation of the observed waveform \eqref{eq:gwf} will directly yield the value of the luminosity distance $d_L$ of the source, implying that binaries emitting GWs can be used as cosmological distance indicators.
For this reason GW sources of this kind are called {\it standard sirens} \cite{Holz:2005df}, in analogy with SNIa which are called standard candles.
If besides the value of $d_L$, obtained from the analysis of the detected GW signal, also the redshift of the GW source (or the one of its hosting galaxy) is measured, then one obtains a data point in the distance-redshift diagram, where the \textit{distance-redshift} relation, defined as
\begin{equation}
	d_L(z) = \frac{c}{H_0}\frac{1+z}{\sqrt{\Omega_k}} \sinh \left[ \sqrt{\Omega_k} \int_0^z \frac{H_0}{H(z')} dz' \right] \,,
	\label{eq:dL_z}
\end{equation}
can be fitted against the observational data.
Here $\Omega_k$ is the relative energy density of spacetime curvature and a FRW universe has been assumed.
$H(z)$ is the Hubble rate whose redshift evolution depends on the specific cosmological model to be tested. 

Thanks to the well-known theoretical understanding of binary inspirals, standard sirens have the advantage over SNIa to be free of unwanted systematic errors and calibration procedures, meaning that they provide a direct measure of the luminosity distance.
Unfortunately determining the redshift of the GW source might be complicated.
The easiest way to measure it is through the observation of an EM counterpart, which however is not guaranteed to be detected or even to be produced.
Creation of an EM counterpart depends on the type of binary considered and on the environment in which the merger occurs.
The current astrophysical models contain many uncertainties due to the lack of enough experimental data.
Moreover a successful detection of an EM counterpart depends also on the sky angular resolution obtained by the GW detector and by the specifics of EM telescopes.
Ideally what is needed is a GW detector able to provide an accurate sky localization to several wide field telescopes able to observe possible EM transients.
In general the situation with GWs is the opposite of the one experienced with EM waves: with standard sirens it is easy to measure $d_L$ (directly from the waveform) but it is difficult to measure $z$ (need an EM counterpart), with standard candles it is easy to measure redshifts (comparing EM spectra) but it is difficult to measure the luminosity distance (need objects of known absolute luminosity).

\section{Using MBHB mergers as standard sirens with LISA}

Any GW signal detected by LISA and produced by a binary astrophysical system, could in principle be used as standard siren.
The expected list includes MBHBs from $10^4$ to $10^7 M_\odot$, stellar BH binaries from 10 to 100 $M_\odot$ and extreme mass ratio inspirals (EMRIs).
Nevertheless only MBHBs are expected to produce a detectable EM counterpart since they are supposed to merge in a gas rich environment and within the LISA frequency band.
They are moreover characterized by a high LISA SNR and abundant at very high redshifts, up to $z \sim 15$, implying that MBHB standard sirens will probe the cosmic expansion at distances SNIa cannot reach.
The following analysis will thus be focused on MBHBs, and it will be aimed at characterizing the number of GW detections with LISA and the fraction of sources for which a EM counterpart will likely be observed by future telescope facilities.
Once these results are obtained, then constraints on cosmological parameters will be forecast.
The procedure employed in \cite{Tamanini:2016zlh} to simulate MBHB merger rates, compute the GW signal detected by LISA and model and observe EM counterparts, will be briefly summarised here.
The reader interested in the details can find them in \cite{Tamanini:2016zlh}.

First of all, one needs to model realistically the expected MBHB sources.
As a starting point the results of semi-analytical simulations of the evolution of the BH masses and spins during the
\begin{wraptable}{r}{.5\textwidth}
	%\begin{table}
	\vspace{-.3cm}
	\begin{center}
	\centerline{\begin{tabular}{|c|c|c|c|c|}
		\hline
		 & SNR   & $\Delta\Omega<$   & EM     & \multirow{2}{*}{\%} \\
		 & $>$ 8 & $10\,{\rm deg}^2$ & count. & \\
		\hline
		\multirow{3}{*}{N2A2M5L6} 
							  & 360 & 35.4 & 28.2 & 7.83 \\
							  & 41.1 & 28.9 & 26.4 & 64.2 \\
							  & 610 & 214. & 40.5 & 6.64 \\
		\hline
		\multirow{3}{*}{N2A5M5L6} 
							  & 683 & 75.2 & 45.8 & 6.71 \\
							  & 41.1 & 35.3 & 31.2 & 75.9 \\
							  & 611 & 385. & 50.0 & 8.18 \\
		\hline
	\end{tabular}}
	\end{center}
	%\captionsetup{width=1.12\textwidth}
	\caption{\it Average values (5 years) of counterpart detections for both N2A2M5L6 and N2A5M5L6 and all three MBHB formation models: popIII, Q3d and Q3nod (respectively from top to bottom in each cell). From left to right the table shows: the LISA detections, the LISA detections with sky location error below 10 deg$^2$, the optical counterparts (observed either with LSST or with SKA+ELT) and their percent fraction with respect to the total number of detection (first column). Results taken from Ref.~\cite{Tamanini:2016zlh}.}
	\label{tab:ss2}
	\vspace{-.3cm}
	%\end{table}
\end{wraptable}
hierarchical galaxy formation and evolution, are used in analogy with the analysis of \cite{PaperI}.
This allows one to predict the rate and redshift distribution of MBHB merger events.
Several variants of the semi-analytical model are produced by considering competing scenarios for the initial conditions of the massive BH population at high redshift -- namely, a ``light-seed'' scenario in which the first massive BHs form from the 
remnants of population III (popIII) stars, and a ``heavy-seed'' one where massive BHs form from the collapse of protogalactic disks -- and for the delays with which massive BHs merge after their host galaxies coalesce (Q3d and Q3nod). 
The simulations produce synthetic catalogues of MBHB merger events, including all information about the MBHBs and their host galaxies.
The parameters of the MBHB systems of each catalogue are then inserted as input into a code that simulates the GW signal induced in LISA by the binaries' inspirals.
Considering all phases of the observed signal (inspiral, merger and ringdown), the code computes the SNR of each merger event and the Fisher matrix of the corresponding waveform parameters, which includes in particular the $1\sigma$ error on the luminosity distance $\Delta d_L$, and on the sky location $\Delta \Omega$.
Among the MBHB merger events, those that have ${\rm SNR}>8$ and $\Delta \Omega <10\,{\rm deg}^2$ are selected.
This guarantees that the events are measured with a directional uncertainty sufficiently small to allow EM telescopes to detect a counterpart, if it is present.

The following step is to select, among these events, those that are likely to provide a detectable EM counterpart.
Several investigations (e.g.~\cite{Armitage:2002uu,Palenzuela:2010nf,Giacomazzo:2012iv}) suggest that optical/radio emission is likely to happen at MBHB mergers in a transient fashion.
In light of these results the counterpart generated by each MBHBs merger is modelled taking into account all the information on the binary and the hosting galaxy.
\begin{wrapfigure}{l}{.5\textwidth}
	\vspace{-.25cm}
	\centering\includegraphics[width=.45\textwidth]{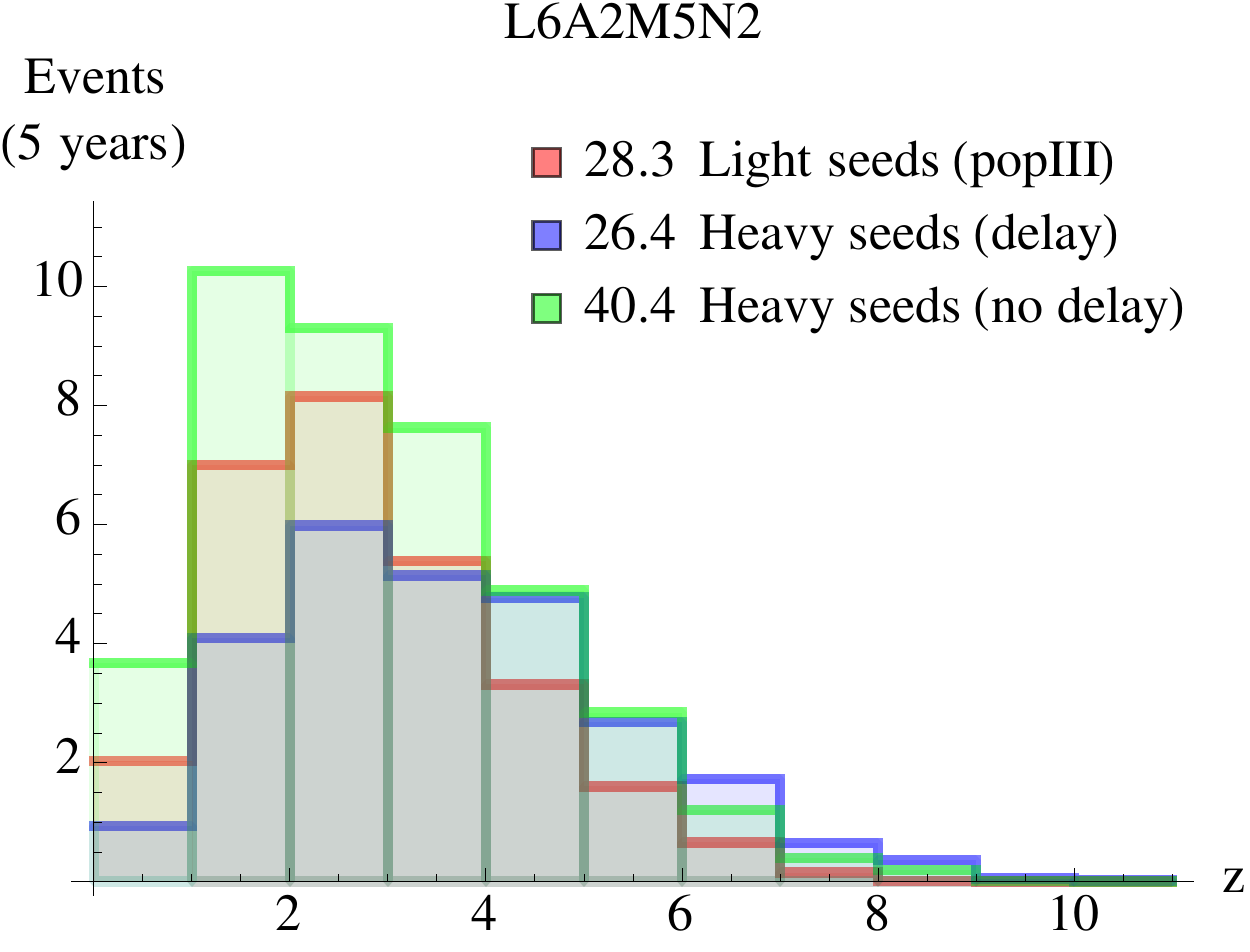}
	\caption{\it Redshift distribution of MBHB standard sirens for N2A2M5L6. From Ref.~\cite{Caprini:2016qxs}.}
	\label{fig:z_distrib_SS}
	\vspace{-.5cm}
\end{wrapfigure}
Generically a quasar-like luminosity flare in the optical and magnetic field induced flare and jet in the radio, are likely to be produced at mergers.
In order to be able to detect these associated EM emissions, ideally optical/radio telescopes should be pointed in the direction of the event prior to merger.
Thus as soon as the event has been localised in the sky with the required precision, telescopes are alerted and pointed in that direction, looking for a distinctive flare occurring at merger.
In the analysis presented here the specifics of future realistic telescopes, such as LSST, SKA and ELT \cite{EMweb}, are considered to observe the counterpart and determine the redshift of the GW source; see \cite{Tamanini:2016zlh} for details.

The LISA design considered in this analysis is labelled with N2A2M5L6 and defined by: armlength of 2 million km, number of active laser links (arms) fixed to six (three), mission duration of five years and the low frequency noise level assumed to be the one tested by LISA pathfinder (N2 in \cite{Tamanini:2016zlh,PaperI}).
This choice is made in agreement with the first results obtained by the LISA pathfinder mission \cite{Armano:2016bkm} and considering a conservative value for the expected interferometer armlength.
Note that, according to \cite{Tamanini:2016zlh,Caprini:2016qxs}, the constraints on cosmological parameters forecast with N2A2M5L6 are basically equivalent to the ones obtained with N2A5M5L6 (same design with 5 million km).
In what follows, unless otherwise specified, all results will refer to N2A2M5L6.

\section{Forecast cosmological constraints}

\begin{wraptable}{r}{.4\textwidth}
\vspace{-.8cm}
\begin{tabular}{|c|c|c|c|}
 \hline
 & $\Delta\Omega_M$ & $\Delta \Omega_\Lambda$ & $\Delta h$ \\
 \hline
 \multirow{3}{*}{$\Lambda$CDM} &
  0.0333 & 0.0333 & 0.0141 \\
& 0.0378 & 0.0378 & 0.0204 \\
& 0.0207 & 0.0207 & 0.0087 \\
 \hline
 \multirow{3}{*}{$k$-$\Lambda$CDM} &
  0.0566 & 0.194 & 0.0370 \\
& 0.0954 & 0.234 & 0.0581 \\
& 0.0287 & 0.105 & 0.0179 \\
 \hline
 & $\Delta w_0$ & $\Delta w_a$ & \\
 \hline
 \multirow{3}{*}{DDE} &
  0.173 & 0.935 & \\
& 0.247 & 1.17 & \\
& 0.108 & 0.582 & \\
\hline
\end{tabular}
\caption{\it 1$\sigma$ errors on cosmological parameters for N2A2M5L6 and all three MBHB models: popIII, Q3d and Q3nod (respectively from top to bottom in each cell).}
\label{tab:errors}
\vspace{-.5cm}
\end{wraptable}
\begin{figure}[b]
	\vspace{-.5cm}
	\centering\includegraphics[width=\textwidth]{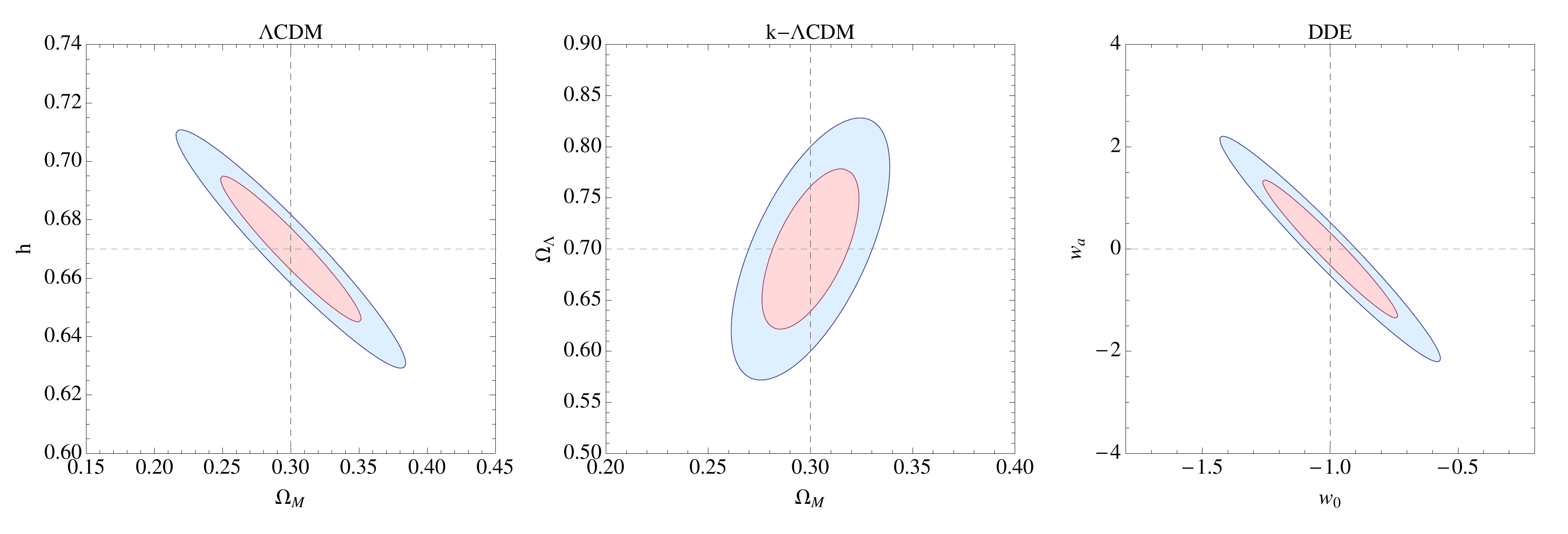}
	\caption{\it 1 (red) and 2$\sigma$ (blue) contour plots in the parameter spaces of $\Lambda$CDM, $k$-$\Lambda$CDM (with $h$ fixed to its fiducial value) and DDE, respectively from left to right. Here only results for N2A2M5L6 with the popIII MBHB model are shown.}
	\label{fig:ellipses}
\end{figure}
The number of MBHB mergers detected by LISA (with an SNR $>8$), the ones with a sufficiently accurate sky localization ($\Delta\Omega< 10\, {\rm deg}^2$) and the ones for which an EM counterpart is observed, are exposed in Table~\ref{tab:ss2}.
In the last column the fractional percentage between the number of standard sirens and the number of MBHB detected by LISA, is reported.
The expected number of MBHB standard sirens is roughly 30 for N2A2M5L6 over the whole five year mission (Table~\ref{tab:ss2} shows also numbers for N2A5M5L6 for comparison).
These numbers are expected to scale linearly if a different mission duration is considered.
The standard sirens distribution in redshift for N2A2M5L6 is given in Fig.~\ref{fig:z_distrib_SS}.
One can immediately notice that the bulk of data will appear in the range $1<z<4$, with a peak at $z \simeq 2$ and tails extending up to $z \simeq 8$.
This implies that LISA will be able to directly probe the expansion of the universe at redshifts not explored by SNIa standard candles which are limited at $z \lesssim 2$.

At this point one can fit these standard sirens data with any cosmological model of interest in order to derive constraints on free parameters.
Here three standard cosmological models will be first considered:
{\bf $\Lambda$CDM}: the standard concordance model with cold dark matter and a cosmological constant (free parameters $\Omega_M, h$);
{\bf $k$-$\Lambda$CDM}: the standard concordance model plus curvature (free parameters $\Omega_M, h, \Omega_\Lambda$);
{\bf DDE}: a dynamical DE model where the DE equation of state is $w (z) = w_0 + w_a z / (z+1)$ and all $\Lambda$CDM parameters are fixed (free parameters $w_0, w_a$).
Fiducial values for these parameters are: $\Omega_M = 0.3$, $\Omega_\Lambda = 0.7$, $h = 0.67$, $w_0 = -1$ and $w_a = 0$.
Each of these models gives rise to a different Hubble rate function $H(z)$ determined by the free parameters of the model.
Once this function is inserted into \eqref{eq:dL_z}, then the theoretical distance-redshift relation can be fitted against the standard sirens data and constraints on the cosmological parameters can be obtained statistically.
The forecast constraints for N2A2M5L6 derived in \cite{Tamanini:2016zlh}, using Fisher matrix techniques, are reported in Table~\ref{tab:errors}.
In Fig.~\ref{fig:ellipses} one and two sigma parameter regions constrained by LISA (N2A2M5L6) for the three cosmological models are shown (for the middle panel, $k$-$\Lambda$CDM model, $h$ has been fixed to its fiducial value).
Note the interesting constraint recovered on $H_0$ for $\Lambda$CDM, which is roughly at the 2\% level, and reduces to 1\% if $\Omega_M$ is fixed to its fiducial value.
This might help in solving the current tension between local and CMB measurements of $H_0$, or at least in providing an independent check.

One can now compare the LISA forecasts with the results obtained by present cosmological probes such as SNIa analyses \cite{Betoule:2014frx} and CMB experiments \cite{Ade:2015xua,Ade:2015rim}.
For $\Lambda$CDM the constraint on $H_0$ obtained by Planck \cite{Ade:2015xua} is only a factor of about two better than the ones forecast for LISA, while the constraint on $\Omega_M$ is better by a factor of about three.
On the other hand the LISA forecast error on $\Omega_M$ once $H_0$ is fixed to its fiducial value, is $\Delta\Omega_M \simeq 0.009$,
\begin{wrapfigure}{r}{.45\textwidth}
	\vspace{-.35cm}
	\centering\includegraphics[width=.45\textwidth]{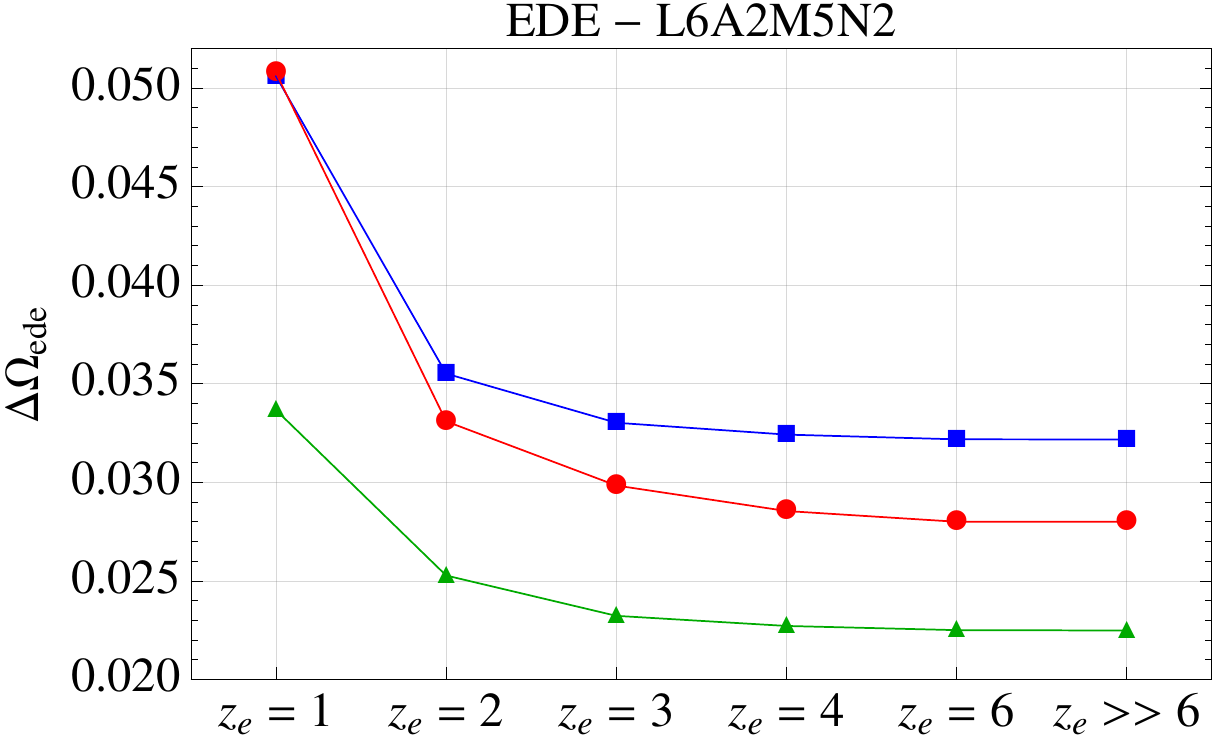}
	\centering\includegraphics[width=.45\textwidth]{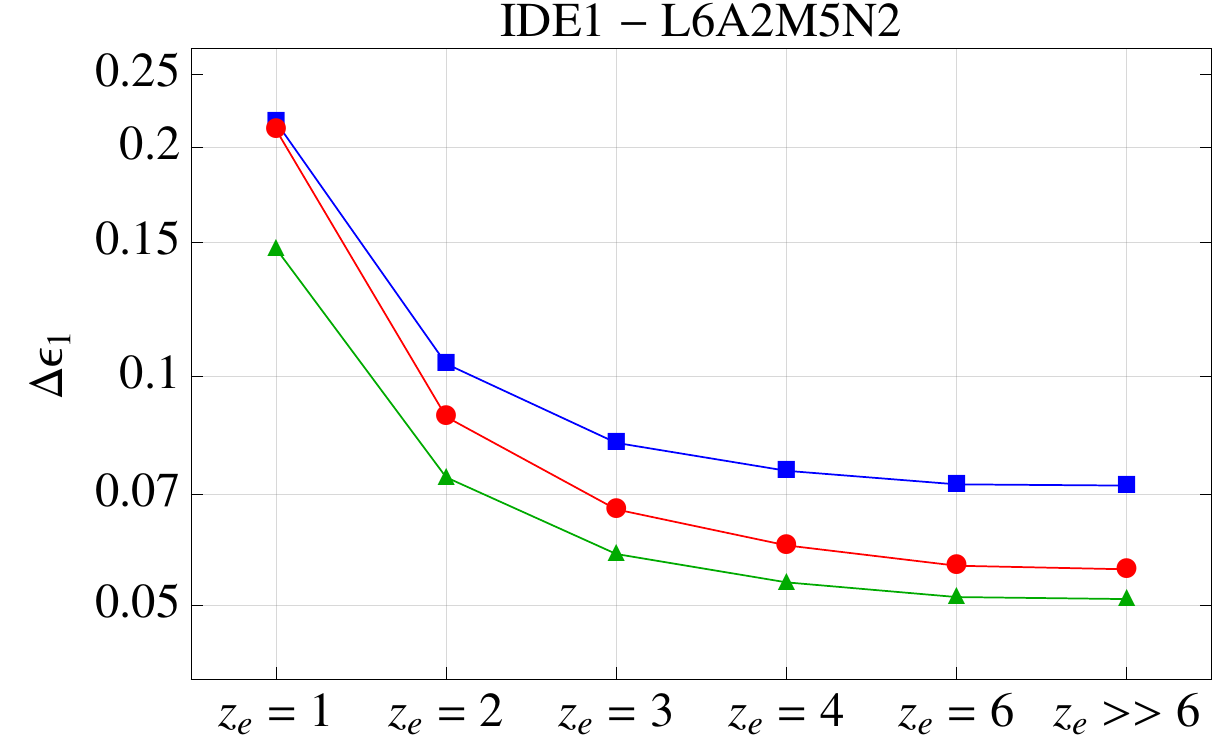}
	\centering\includegraphics[width=.45\textwidth]{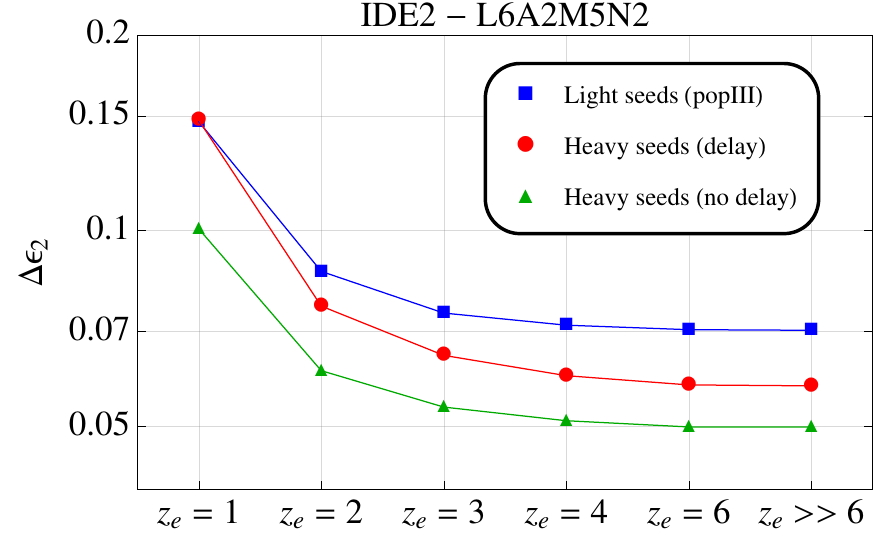}
	\caption{\it 1$\sigma$ constraints from Ref.~\cite{Caprini:2016qxs} on alternative cosmological models.}
	\label{fig:alt_models}
	\vspace{-.5cm}
\end{wrapfigure}
which is a factor of two better than the one obtained by present SNIa \cite{Betoule:2014frx} (SNIa cannot measure $H_0$).
Finally the constraints for DDE are comparable to the ones obtained with all present probes (CMB+SNIa+BAO) combined \cite{Betoule:2014frx}.
These numbers show the potential of the LISA mission: the constraints obtained by a single spaceborne GW probe and only with one population of sources (MBHBs), are roughly at the same level of the present constraints obtained by all combined cosmological probes.
However one must keep in mind that future conventional EM probes, such as Euclid \cite{Laureijs:2011gra}, will be able to obtain much better constraints than the present ones and thus to be more competitive than LISA.
Nevertheless the results that will be obtained with LISA, and with GW experiments in general, will constitute a new way to test the expansion of the universe.

The same analysis can be performed for other cosmological models.
Following \cite{Caprini:2016qxs} the results for two simple alternative models will be presented.
They constitute one-parameter extensions of $\Lambda$CDM and are defined as:
{\bf EDE}: an early DE model where the relative energy density of DE remains non negligible at early times \cite{Doran:2006kp,Pettorino:2013ia} (free extra parameter $\Omega_{\rm ede}$);
{\bf IDE}: a model where DE interacts with DM by an energy exchange proportional to the energy density of either DM (IDE1: free extra parameters $\epsilon_1$) or DE (IDE2: free extra parameters $\epsilon_2$), see \cite{Wang:2016lxa} for a recent review.
Both these models are assumed to effectively describe the cosmic expansion up to some reference redshift $z_e$ after which the standard $\Lambda$CDM dynamics is recovered (details can be found in \cite{Caprini:2016qxs}).
This is done in order to test the ability of LISA to probe models where the deviations from $\Lambda$CDM happen only at late cosmological times.
There are in fact for example some observational indications that an interaction in the dark sector might be present at late times, while being negligible at early times \cite{Salvatelli:2014zta}, where CMB experiments provide stringent constraints.
Assuming an exact prior on the $\Lambda$CDM parameters, the 1$\sigma$ error forecasts for LISA on the extra alternative parameters are shown in Fig.~\ref{fig:alt_models} for different values of $z_e$.
For each model the accuracy on the extra parameter reaches few $10^{-2}$, worsening as $z_e$ decreases and remaining constant for $z_e \gtrsim 6$.
If the deviations from $\Lambda$CDM are relevant up to redshifts larger than about 10, then these constraints are worse, or at most comparable to, present CMB constraints \cite{Caprini:2016qxs}.
However if these deviations are effectively important only after $z_e \simeq 10$, then CMB experiments cannot generally impose the same bounds they find for $z_e$ much bigger than 10; see \cite{Ade:2015rim}.
This shows the main strength of LISA as a cosmological probe: it will test the cosmic expansion in the range $1 < z < 8$ where any deviation from $\Lambda$CDM will be directly constrained, irrespectively of it being negligible or not at earlier times.

\section{Conclusions and future prospects}

\begin{wrapfigure}{r}{.5\textwidth}
	\vspace{-.8cm}
  \centering\includegraphics[width=.5\textwidth]{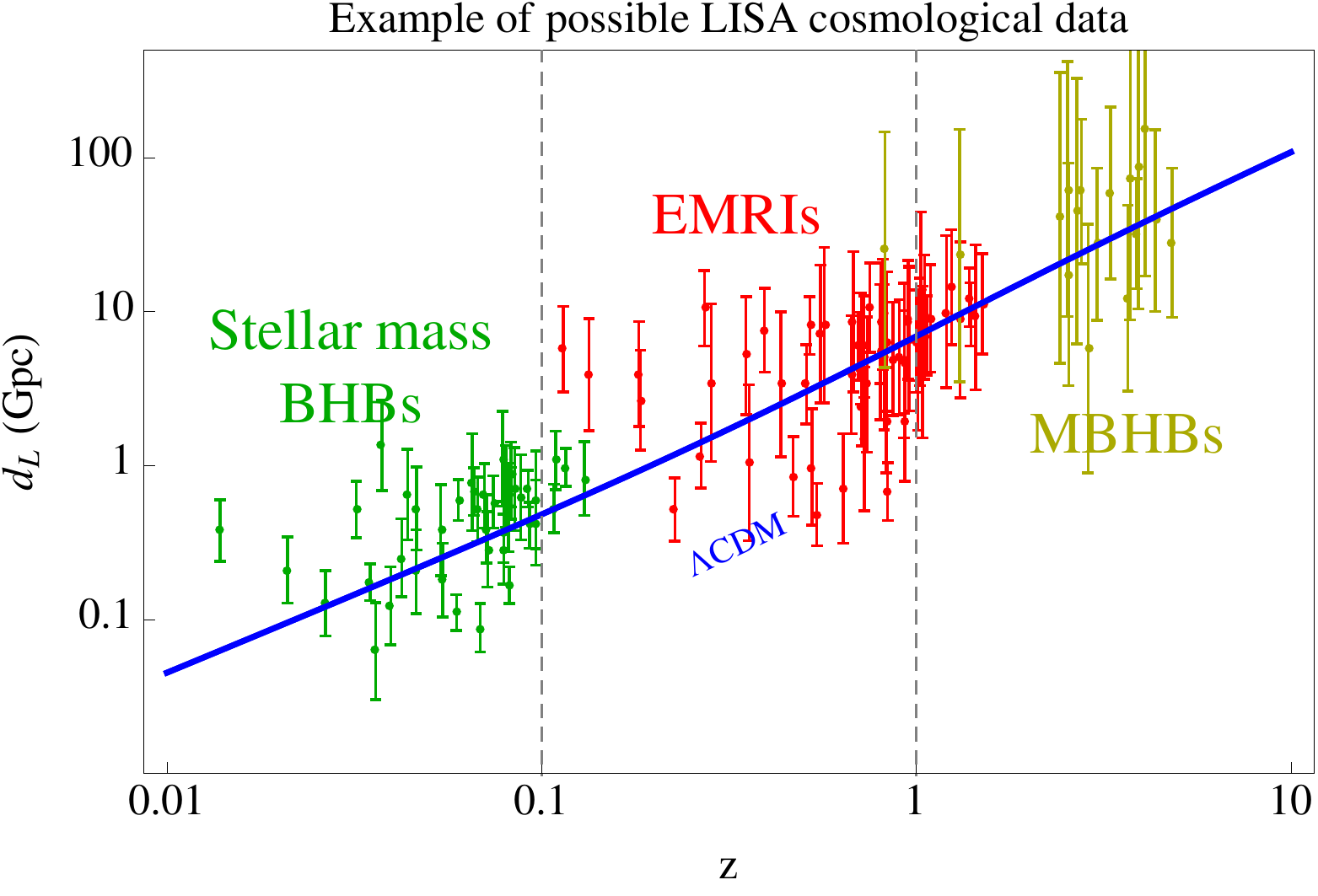}
  \caption{\it Cosmology at all redshifts with LISA.}
  \label{fig:all_z}
\end{wrapfigure}
GW standard sirens will in general offer a new independent way, complementary to common EM observations, to probe the cosmic expansion.
Even if rare, GW standard sirens will complement, and increase confidence in, other cosmic distance indicators, in particular SNIa standard candles.
MBHB mergers will constitute excellent standard sirens for LISA, allowing for systematic and calibration free distance indicators up to very high redshifts ($z \sim 8$).
The biggest challenge remains the identification of an EM counterpart needed to measure the redshift of the GW source.
Using a simple model of EM emission at merger and assuming the specifics of future telescopes, in \cite{Tamanini:2016zlh} it was shown that LISA (N2A2M5L6) will be able to detect around 30 MBHB standard sirens over an observational period of five years.
With these numbers the constraints forecast for standard and alternative cosmological parameters are roughly at the level of the present cosmological probes combined, but will likely be overcome by future observations.
Nevertheless LISA will directly explore the expansion of the universe in a redshift range not accessible by other cosmic rulers, such as SNIa.
For this reason the LISA mission can be used to obtain new and independent observational data to constrain standard and especially alternative models at very high redshift.
On the other hand, MBHBs are not the only GW sources that LISA can use as standard sirens: stellar mass BHBs and EMRIs can also be used as distance indicators by LISA. 
Although no counterpart is expected for these sources, other (statistical) methods can be applied to overcome the lack of redshift information \cite{Kyutoku:2016zxn}.
Interestingly stellar mass BHBs are usually detected in the range $z \lesssim 0.1$, while EMRIs appear at redshifts $0.1 \lesssim z \lesssim 1$, where the acceleration of the universe can be better tested.
Considering the fact that MBHB standard sirens provide data in the $1 \lesssim z \lesssim 10$ range, this implies that LISA will become a unique cosmic probe able to test all redshift ranges up to $z \sim 10$, as qualitatively depicted in Fig.~\ref{fig:all_z}.

\section*{References}
\bibliography{bibfile,COSMO_paper}

\end{document}